\documentclass[aps,superscriptaddress,twocolumn,noshowpacs,noshowkeys]{revtex4}

\usepackage{dcolumn}
\usepackage{amsmath}
\usepackage{bm}
\usepackage{graphicx}
\usepackage{graphics}

\begin{document}

\title{Extracting the Dynamic Magnetic Contrast in Time-Resolved X-ray Transmission Microscopy}

\author{T. Schaffers}
\affiliation{Institute of Semiconductor and Solid State Physics, Johannes Kepler University Linz, 4040 Linz, Austria}
\author{T. Feggeler}
\affiliation{Faculty of Physics and Center for Nanointegration Duisburg-Essen (CENIDE), University of Duisburg-Essen, 47057 Duisburg, Germany}
\author{S. Pile}
\affiliation{Institute of Semiconductor and Solid State Physics, Johannes Kepler University Linz, 4040 Linz, Austria}
\author{R. Meckenstock}
\affiliation{Faculty of Physics and Center for Nanointegration Duisburg-Essen (CENIDE), University of Duisburg-Essen, 47057 Duisburg, Germany}
\author{M. Buchner}
\affiliation{Institute of Semiconductor and Solid State Physics, Johannes Kepler University Linz, 4040 Linz, Austria}
\author{D. Spoddig}
\affiliation{Faculty of Physics and Center for Nanointegration Duisburg-Essen (CENIDE), University of Duisburg-Essen, 47057 Duisburg, Germany}
\author{V. Ney}
\affiliation{Institute of Semiconductor and Solid State Physics, Johannes Kepler University Linz, 4040 Linz, Austria}
\author{M. Farle}
\author{H. Wende}
\affiliation{Faculty of Physics and Center for Nanointegration Duisburg-Essen (CENIDE), University of Duisburg-Essen, 47057 Duisburg, Germany}
\author{S. Wintz}
\affiliation{Paul Scherrer Institut, 5232 Villigen PSI, Switzerland}
\affiliation{Helmholtz-Zentrum Dresden-Rossendorf, 01328 Dresden, Germany}
\author{M. Weigand}
\affiliation{Max-Planck-Institut f{\"u}r Intelligente Systeme, 70569 Stuttgart, Germany}
\author{H. Ohldag}
\affiliation{Stanford Synchrotron Radiation Laboratory, SLAC National Accelerator Laboratory, Menlo Park, California 94025, USA}
\author{K. Ollefs}
\affiliation{Faculty of Physics and Center for Nanointegration Duisburg-Essen (CENIDE), University of Duisburg-Essen, 47057 Duisburg, Germany}
\author{A. Ney}
\email{andreas.ney@jku.at; Phone: +43-732-2468-9642; FAX: -9696}  
\affiliation{Institute of Semiconductor and Solid State Physics, Johannes Kepler University Linz, 4040 Linz, Austria}

\begin{abstract}

Using a time-resolved detection scheme in scanning transmission X-ray microscopy (STXM) we measured element resolved ferromagnetic resonance (FMR) at microwave frequencies up to 10\,GHz and a spatial resolution down to 20\,nm at two different synchrotrons. We present different methods to separate the contribution of the background from the dynamic magnetic contrast based on the X-ray magnetic circular dichroism (XMCD) effect. The relative phase between the GHz microwave excitation and the X-ray pulses generated by the synchrotron, as well as the opening angle of the precession at FMR can be quantified. A detailed analysis for homogeneous and inhomogeneous magnetic excitations demonstrates that the dynamic contrast indeed behaves as the usual XMCD effect. The dynamic magnetic contrast in time-resolved STXM has the potential be a powerful tool to study the linear and non-linear magnetic excitations in magnetic micro- and nano-structures with unique spatial-temporal resolution in combination with element selectivity.

\end{abstract}

\maketitle

\section{Introduction}

\noindent In spintronics and magnonics it is imporant to understand the magnetization dynamics on the micro- and nano-scale e.\,g. to be able to control the propagation of spin waves. A well-established technique to measure the dynamic magnetic behavior of a system is ferromagnetic resonance (FMR). Yet classical resonator based FMR measurements are not able to detect single micro- or nano-sized objects due to their detection limit of around $10^{11}$ spins \cite{POO97}. This sensitivity limit has been overcome in recent years by the development of lithographically fabricated micro-resonators \cite{NAR05} which are capable of measuring down to $10^{6}$ spins \cite{BAN11}, corresponding to a single Fe-nanocube with dimensions of $30\times30\times30\,nm^{3}$. Due to the lack of spatial resolution below the diameter of the micro-resonator of typically a few tens of microns it is impossible to separate the FMR signal of a single nano-particle from the resonance signal of the whole ensemble during the homogeneous excitation of the micro-resonator cavity.\\
To facilitate spatial resolution other measurement techniques have been combined with FMR excitation in order to measure a single nano-sized object in an ensemble. These measurement techniques include but are not limited to: magneto optic Kerr effect (MOKE) \cite{ROS02}, Brillouin light scattering (BLS) \cite{DEM01}, magnetic force microscopy (MFM) \cite{VOL04}, scanning thermal microscopy (SThM) \cite{SCH17}, scanning electron microscopy with polarization analysis (SEMPA) \cite{SCH18}, and X-ray photoemission electron microscopy (X-PEEM) \cite{CHE12}. For most of these measurement techniques it is not possible to measure with element selectivity (MOKE, BLS, MFM, SThM and SEMPA), while other measurement techniques like X-PEEM can only probe the surface of the sample with element selectivity. In recent years the X-ray magnetic circular dichroism (XMCD) \cite{SCH86,STO95,DUR09} effect has been combined with FMR in order to probe the dynamic magnetic excitation, the so called X-ray detected ferromagnetic resonance (XFMR) \cite{OMS15}, utilizing the element selectivity of the X-rays. A spatial resolution of down to 20\,nm can be achieved by using a scanning transmission X-ray microscope (STXM); however, initially the time-resolution was restricted to below 1~GHz \cite{PUZ05}. By combining the micro-resonator FMR with STXM (STXM-FMR) within a synchronization scheme for the exciting microwaves and the probing X-ray photons of the synchrotron it is possible to detect FMR with a high temporal (ps-regime) as well as spatial resolution (nm regime) \cite{WEI14,Bon15,SCH17}. Combining these features STXM-FMR measurements bare the potential to significantly deepen our understanding of, e.\,g., magnetic domain wall dynamics \cite{STE13}, spin wave emitters \cite{WIN16}, the dynamic magnetization of ferromagnetic heterosystems containing different chemical elements \cite{FEG19} as well as non-ferromagnets with induced magnetization\cite{KUK15}.\\
In order to be able to draw valid conclusions from the dynamic magnetic contrast in STXM-FMR, it is necessary to perform a range of control-experiments in the first place as well as testing the robustness of the evaluation of the raw data to establish that STXM-FMR indeed provides significant information about the dynamic magnetic behavior of a given magnetic specimen based on the XMCD effect. In this paper a range of control experiments will be presented as well as a detailed analysis of the separation of the true magnetic contrast from background effects. The obtained results allow to reliably image homogeneous and inhomogeneous magnetic excitations in magnetic micro-stuctures with very high spatio-temporal resolution. Furthermore, it is possible to obtain quantitative information about the local precession angle in FMR and its relative phase within a given STXM-FMR experiment.

\section{Experimental Details}

\begin{figure}[h!]
\center
\resizebox{1\columnwidth}{!}{\includegraphics{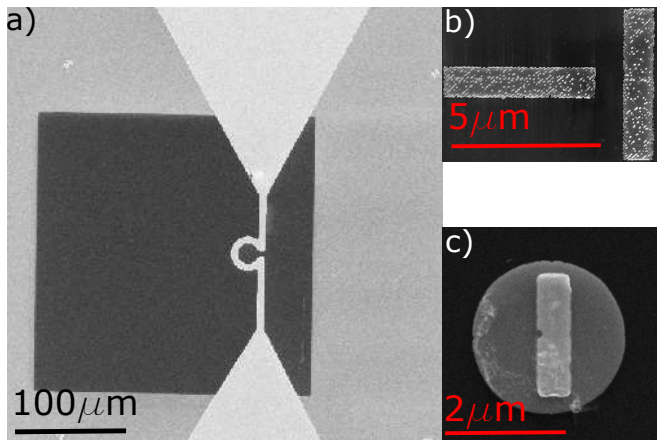}}
\caption{Scanning electron microscope (SEM) image of the strip-line resonator on top of a SiN-membrane. For this work two different sample systems were chosen. In b) two perpendicular Py stripes ("T"-sample) are shown, while in c) the Py-Co disk stripe sample can be seen.}
\label{Fig_sa}
\end{figure}

\noindent The magnetic specimen is placed inside a micro-resonator and microwaves are used to excite the FMR. The micro-resonator is fabricated on a 200\,nm thick, 250$\times$250\,$\mu$m$^{2}$ large silicon nitride membrane suspended by a 5$\times$10\,mm$^{2}$ silicon frame of high resistivity. In a first step the magnetic specimen is fabricated on the SiN-membrane using electron beam lithography (EBL). Two different designs for the magnetic specimen were made. The first one consists of two perpendicular permalloy (Py) stripes with dimensions of 5$\times$1$\times$0.03\,$\mu$m$^{3}$ (see Fig.\,\ref{Fig_sa}\,b)), which are deposited using magnetron sputtering at room temperature and capped with aluminum. The second sample system is a combination of a Py disk with a Co stripe. For this in a first step a Py disk with a diameter of 2.6\,$\mu$m and a thickness of 30\,nm is fabricated. With a second EBL step a Co stripe with lateral dimensions of 2$\times$0.6\,$\mu$m$^{2}$ and a thickness of 30\,nm is placed on top of the Py disk (see Fig.\,\ref{Fig_sa}\,c)). Finally, the micro-resonator is patterned around the magnetic specimen using optical lithography (OL), leaving the sample inside the $\Omega$ shaped resonator loop in Fig.\,\ref{Fig_sa}\,a). The gold used to produce the micro-resonator has a thickness of 600\,nm and an additional 5\,nm of titanium is used as an adhesion layer. Both materials are deposited by thermal evaporation.\\
To measure the STXM-FMR at the Stanford Synchrotron Radiation Lightsource (SSRL), beamline 13.1, the FMR excitation needs to be synchronized to the bunch freuqency of the synchrotron, thus enabling us to measure the FMR precession with time resolution. By utilizing the STXM it is possible to measure with a spatial resolution, 35\,nm at SSRL and 20\,nm at the MAXYMUS beamline at BESSY II, achieved by focusing the X-rays onto the magnetic specimen using a zone-plate. The stroboscopic time resolution for the STXM-FMR measurement is achieved by phase locking the GHz microwave frequency to the 476.315 MHz bunch frequency of the SSRL synchrotron. Furthermore, a PIN-diode was installed to switch the microwave on and off with the synchrotron revolving frequency of 1.28\,MHz. This comparison of X-ray transmission detected with and without applied microwave power allows to detect very small changes in the x-ray transmission as a result of the magnetization precession. A fast avalanche photo diode (APD) detects the transmitted X-ray photons behind the sample. The APD signal is finally stored in 12 different channels. Each of these channels corresponds to the signal of a specific group of X-ray pulses. The first 6 channels are used for the APD signal of the transmitted X-rays with applied microwave, while the second 6 channels are used to measure the X-ray transmission without applied microwave. For the first 6 channels the magnetization inside the sample is precessing, while the magnetization is static for the second 6 channels. Each of these channels corresponds to a specific relative phase of the FMR precession with respect to the microwave excitation. The latter non-precessing channels are crucial to eliminate the influence of the filling pattern of the bunch train of electron buckets on the resulting STXM images. The 6 different channels correspond to 6 specific bunches which are phase shifted each by 60$^\circ$, with respect to the microwave frequency of up to 9.6\,GHz. Therefore, the 6 phases correspond to time resolved snap-shot images which are separated by 17.4\,ps, and comprise one full precession cycle of the magnetization. One should note, that each X-ray flash has a pulse duration of 50\,ps, which fundamentally limits the attainable temporal resolution. Additional information regarding the synchronization scheme and the X-ray microscope can be found in \cite{SCH17, Bon15}.\\\\
A similar approach for measuring the dynamic magnetization in an FMR experiment with spatial and temporal resolution has been implemented at the Maxymus endstation at BESSY II \cite{WEI14}. There are two moderate differences with respect to the SSRL experiment: One is that the BESSYII operation frequency is appx. 500 MHz, corresponding to a repetition period of the probing x-ray flashes of 2 ns. Secondly, the signal is recorded only for the microwave on (precessing) case. Therefore, on the one hand side, it is not necessary to excite the sample at direct higher harmonic frequencies of the synchrotron and thus the exciting frequencies can be chosen more freely to f=500MHz*M/N, depending on the number of detection channels used (N) and a selectable integer multiplier M.\cite{WIN16} Here N for most cases is also equal to the number of simultaneously acquired excitation phases (not limited to 6). Since the not-precessing magnetization (microwave off) cannot be used as a baseline for comparison, on the other hand, for normalization purposes only the transmitted intensity ratio of each channel I(t)/ < I >t with respect to the temporal average state can be evaluated in order to extract the dynamic magnetic contrast originating from the precession of magnetization

\section{Contrast mechanism}

For a better understanding of the measured STXM-FMR data, we briefly discuss the underlying physical effect which yields the dynamic magnetic contrast images.

\subsection{X-ray absorption}

The transmission of electromagnetic radiation through any material is described by Beer-Lambert`s law \cite{SWI62}. In a STXM the transmitted X-ray intensity $I$ is detected. This transmitted intensity is comprised of the X-ray absorption (XA) coefficient of the entire sample (magnetic specimen and SiN-membrane). Tuning the photon energy to any characteristic core-level excitation resulting in the well-known element selectivity of XA measurements. However, the above mentioned law only considers a single layer system. In an STXM-FMR experiment the sample consists at least of a two layer system since any magnetic specimen is suspended on a SiN-membrane through which the X-rays need to be transmitted as well. In order to include this second layer Beer-Lambert`s law needs to be modified \cite{SWI62}:

\begin{equation}
I_{s/m}=I_{0}e^{-(\mu_{s} t_{s}+\mu_{m}t_{m})}
\label{EQ_BLG2L}
\end{equation}

where $t_{s}, t_{m}$ are the thicknesses and $\mu_{s}, \mu_{m}$ are the absorption coefficients of the magnetic specimen and SiN-membrane respectively, and $I_0$ is the incoming intensity. In any sample one can find areas where the X-rays only transmit through the SiN-membrane while in other regions the X-rays are transmitted through the SiN-membrane plus the magnetic specimen, which can also consist of more than one layer which would be added to the exponent in Eq.\,\ref{EQ_BLG2L}. Therefore, from a single STXM-FMR image one can separate the dynamic magnetic contrast from the background transmission of the SiN-membrane by defining respective regions of interest (RoI) from the time-averaged z-contrast images like in Fig.\,\ref{Fig_sa}\,c). 

\section{Analysis of STXM-FMR measurements}

In the light of the preceding discussion, evaluation methods for the extraction of quantitative information from the STXM-FMR data will be presented, with special attention to how to extract the dynamic contribution of the magnetic specimen. Additionally it is possible to quantify the opening angle of the magnetization precession in FMR directly from the change in absorption coefficient during a full precession cycle.  

\subsection{Raw data treatment}
\begin{figure}[h!]
\center
\resizebox{1\columnwidth}{!}{\includegraphics{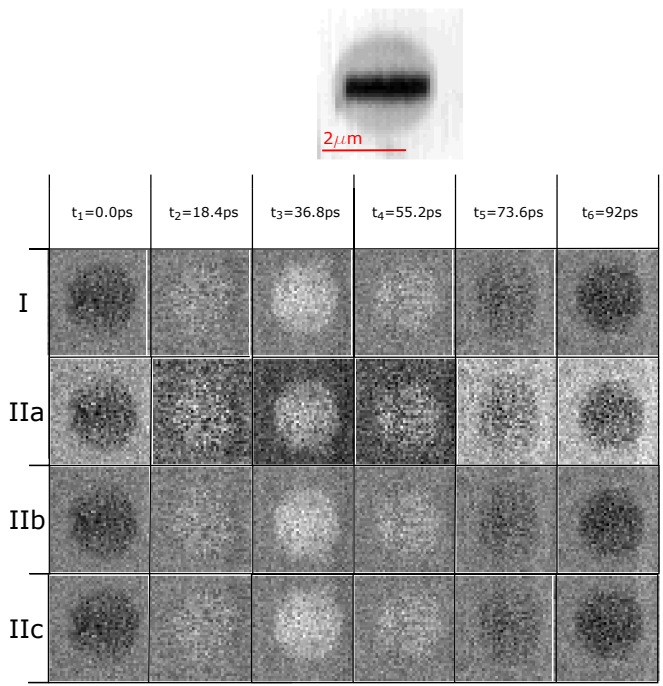}}
\caption{The top shows the chemical contrast of the disk stripe sample measured at the Ni-$L_{3}$-edge. Representation of the different evaluation methods for the 6 phases of the magnetization precession: In I the absorption coefficient difference is shown obtained the background corrected microwave on and off measurement. IIa is the ratio between the not background corrected microwave on and off measurement. Applying a background correction to IIa the images labelled IIb are generated. The images labelled IIc show the absorption coefficient change between microwave power on and off for the different phases.}
\label{Fig_BGcorr_notBGcorr}
\end{figure}
\begin{figure}[h!]
\center
\resizebox{1\columnwidth}{!}{\includegraphics{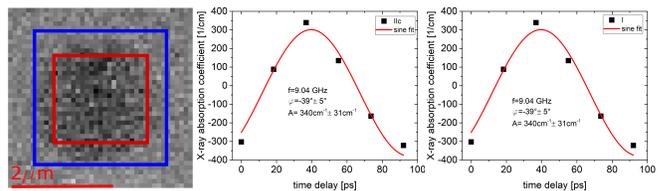}} 
\caption{By averaging over the sample in the six different phases for evaluation method IIc (left side) and I (right side) from Fig.\,\ref{Fig_BGcorr_notBGcorr} one can obtain the shown curves. Both were fitted with a sine fit due to the sinusoidal behavior of the exciting microwave. The frequency of this sinusoidal was given by the microwave frequency applied to the system which in this case was 9.04\,GHz.}
\label{Fig_VGL_ln(ratio)_neweval}
\end{figure}

\noindent To eliminate the second absorption coefficient $\mu_2$ in equation \ref{EQ_BLG2L} the raw data needs to be corrected by the SiN-membrane background. Thus the absorption coefficient of the magnetic specimen alone can be investigated. For that we average the transmission signal over the area of only the SiN-membrane for each of the 12 images (6 phases with microwave on $I^{on}_{m}$ and 6 phases with microwave off $I^{off}_{m}$) separately. Each individual image is then divided by its respective averaged transmission value of the SiN-membrane. The resulting transmission $I^{on}$ and $I^{off}$ then contains exclusively the information about the absorption coefficient $\mu_1$ of the magnetic specimen:
\begin{equation}
I^{on}_s=\dfrac{I^{on}_{s/m}}{I^{on}_{m}}=e^{\mu^{on}_s \cdot t_s} \qquad I^{off}_s=\dfrac{I^{off}_{s/m}}{I^{off}_{m}}=e^{\mu^{off}_s \cdot t_s}
\end{equation} 
 Note, that this also eliminates the dependence on the individual incoming intensity $I_0$ for each phase. Subsequently, the dynamic magnetic contrast is derived by taking the natural logarithm of the ratio of the precessing (microwave on) versus non-precessing (microwave off) case to obtain $\Delta \mu$ corresponding to the difference in absorption coefficient equivalent to the usual definition of the XMCD effect: 
\begin{equation}
ln(\dfrac{I^{on}_s}{I^{off}_s})=(\mu^{off}_s-\mu^{on}_s)\cdot t = \Delta \mu \cdot t
\label{EQ_new_eval}
\end{equation}
where $t$ is the thickness of the magnetic specimen. The resulting dynamic magnetic contrast $\Delta \mu \cdot t$ of the Py disk recorded at the Ni $L_3$-edge at 9.04\,GHz is shown for all six phases in Fig.\,\ref{Fig_BGcorr_notBGcorr}, row I. It is clearly visible that only the contrast of the Py disk reverses during a full measurement cycle representing the perpendicular component of the high-frequency magnetization, while the background stays constant.\\
However, one can change the sequence of extracting $\Delta \mu \cdot t$ and take a closer look at each individual step. First, the ratio of microwave on and microwave off is taken and all 6 phases are displayed in Fig.\,\ref{Fig_BGcorr_notBGcorr} row IIa. It is obvious, that the background corresponding to the SiN-membrane oscillates as well, which will be discussed further below. In a second step, the influence of the oscillating background is corrected as mentioned before by dividing each phase with the respective averaged transmission of the SiN-membrane. The resulting 6 phases are shown in Fig.\ \ref{Fig_BGcorr_notBGcorr}, row IIb and already compare well with the full analysis of I revealing no visible background oscillation.\\
However, the images of IIb do not directly reflect the numerical values of $\Delta \mu \cdot t$. Taking the natural logarithm of IIb one obtains the 6 phases as shown in Fig.\,\ref{Fig_BGcorr_notBGcorr}, row IIc. A direct comparison between IIb and IIc reveals, that the qualitative behavior of the dynamic magnetic contrast is identical. However, only IIc shall be mathematically equivalent to the full analysis in I. Both evaluations depend on the selection of the RoI from which the background of the SiN-membrane is derived.\\
To verify if the sequence of the evaluation steps indeed yield the same results, the quantitative outcome of methods I and IIc are compared in Fig.\,\ref{Fig_VGL_ln(ratio)_neweval}. In the STXM-FMR image the area outside the blue box defines the RoI used for determining the background of the SiN-membrane. The red box indicates the RoI which is used for determining the average $\Delta \mu \cdot t$ of the magnetic specimen. To derive the absorption coefficient $\Delta \mu$ at Ni $L_3$-edge the resulting averaged value has to be divided by the effective thickness $t=24$\,nm, since the Py film is 30\,nm thick and contains 80\% nickel, note that the non-resonant XA of the iron can be excluded due to the ratio between the measurements with and without applied microwave power. The two panels show the averaged values (symbols) of the 6 phases for method I (right) and IIc (left) reflecting the dynamic magnetic contrast of the homogeneous excitation, i.\,e., uniform mode of the Py disk. The sine fits (solid lines) are done for the fixed frequency of the exciting microwave of 9.04\,GHz while amplitude $A$ and phase $\varphi$ are fitting parameters. Indeed, both methods reveal identical numerical values for $A = \Delta \mu$ and $\varphi$ of $(340 \pm 31)$\,cm$^{-1}$ and $-39^{\circ}\pm5^{\circ}$, respectively. The quantitative numerical values will be discussed in the following.

\subsection{Precession angle}
\noindent
The first quantity which can be extracted from a STXM-FMR experiment is the amplitude $A$ corresponding to the dynamic magnetic contrast $\Delta \mu$. For a known thickness $t$ of the magnetic specimen one can extract the opening angle $\theta$ of the precessing magnetization. Other than for the phase $\varphi$, the amplitude $A$ and thus $\Delta \mu$ can be compared between different samples. For that a usual XMCD experiment is carried out on a specimen of known thickness $d$ where $\Delta \mu_{XMCD}$ is derived as the difference in absorption with the magnetization fully parallel and antiparallel to the $k$ vector of the X-rays, yielding $\Delta \mu_{abs}=\Delta \mu_{XMCD} / d$. One should keep in mind that in an XMCD experiment the magnetization is fully reversed while in the STXM-FMR measurement microwave off corresponds to the fully perpendicular case. Therefore, $2A$ has to be taken when comparing with $\Delta \mu_{abs}$. In addition, from geometrical considerations this method yields the full opening angle of the precession cone corresponding to $2\theta$, therefore yielding:
\begin{equation}
sin(2\theta)=\dfrac{2A}{\Delta \mu_{abs}}
\end{equation}
Here $2A$ is $(680 \pm 31)$cm$^{-1}$ and $\Delta \mu_{abs} \approx 200000$\,cm$^{-1}$, which yields an opening angle of $\theta=0.10^{\circ}\pm 0.01^{\circ}$. As already pointed out before \cite{Bon15}, one has to consider the effect of the pulse length of the X-rays on the measured intensity. A pulse length of 50\,ps yields a reduction factor of 1.5 for this measurement at 9.04\,GHz, due to the averaging over part of the dynamic magnetic response of the system. Therefore, the actual opening angle for this FMR measurement is $\theta=0.15^{\circ}\pm 0.02^\circ$ which is of the same order as the previously reported opening angle of $0.1^{\circ}$ for a Co-stripe \cite{Bon15}. It has to be taken into consideration that the obtained opening angle of the FMR is only the out-of-plane angle, which in turn can differ from the in plane angle due to the magnetic anisotropy of the thin film sample.

\subsection{Origin and influence of the background signal}
\begin{figure}[h!]
\center
\resizebox{1\columnwidth}{!}{\includegraphics{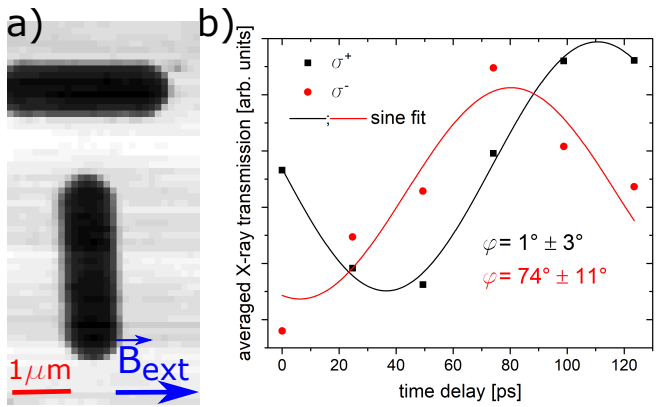}}
\caption{a) shows the chemical contrast images of the measured sample, while the direction of the external magnetic field ($B_{ext}$) is indicated by the blue arrow. b) shows the averaged transmission intensity of the stripe parallel to the external magnetic field for two different X-ray polarizations $\sigma^+$ and $\sigma^-$.}
\label{Fig_no_contrast_rev}
\end{figure}

Due to the background oscillation observed in Fig.\,\ref{Fig_BGcorr_notBGcorr} IIa it is possible that small changes in the dynamic magnetization can not be measured properly. In Fig.\ref{Fig_no_contrast_rev} a measurement on a Py "T"-sample is shown, where the polarization of the X-ray photons is switched from $\sigma^+$ to $\sigma^-$.
\noindent Fig.\,\ref{Fig_no_contrast_rev}\,a) shows the chemical contrast image of the measured sample. The direction of the external magnetic field is indicated by the blue arrow. In Fig.\,\ref{Fig_no_contrast_rev}\,b) the averaged transmission intensity of the stripe parallel to the external magnetic field is shown for the two X-ray polarizations $\sigma^+$ (black curve) and $\sigma^-$ (red curve). The resulting relative phases for the individual measurements are $1^\circ \pm 3^\circ$ for the measurement using $\sigma^+$ and $74^\circ \pm 11^\circ$ for the measurement using $\sigma^-$. As can be seen the phase of the STXM-FMR measurement does not change by 180$^\circ$ as it would be expected for the XMCD effect, when switching the polarization of the incoming X-rays. To understand the reason of this non reversal of the dynamic magnetization the background signal observed in Fig.\,\ref{Fig_BGcorr_notBGcorr} IIa needs to be discussed.\\
The output signal of the avalanche photo diode is amplified by a factor of 1000 (60\,dB) to be detected. Therefore, it is very sensitive to issues with the pre-amplification. The cables inside the STXM (power supply for the APD and signal output of the APD) can act as antennas for standing waves generated by the microwave excitation of the sample. This can cause false positives/negatives depending of the phase of the microwave with regard to the photon arrival time, which can be misinterpreted as bulk (low spatial frequency) dynamics. This is an issue since microvolts of induced voltage by the microwave can be amplified to a "photon" level in the signal output of the APD. \\
Additionally common detection methods can only detect one photon per bunch. Multi photon events only register as single events. This creates a non-linear detector response that gets more pronounced for higher count rates, and can interfere with normalization of dynamic contrast when imaging samples with big static contrast. While the signal in dark areas (magnetic specimen) is linear, the signal in bright areas (SiN-membrane) is compressed, thus the normalization algorithms that work by averaging obtain a skewed response that can create false dynamic contrast proportional to the static contrast.\\
However, if the dynamic contrast reverses when changing the polarization is switched, it can be concluded that the observed signal is a consequence of the dynamic magnetic response of the system to the external excitation generated by the microwave. In order to do so the phase between the microwave excitation and the X-ray pulses can be obtained from the STXM-FMR measurements.

\subsection{Absolute vs. relative phase}
The absolute phase should be measured between the precessing magnetization and the arrival of the X-ray pulse. However, this is complicated due to several issues. First, as in any resonance experiment, there is a phase difference between the microwave excitation and the precessing magnetization. Second, the phase of the X-ray pulse cannot be determined directly since only the driving frequency of the rf-cavity of the storage ring is accessible. Therefore, the travel time of the electron bunches from the cavity to the undulator as well as the travel time of the X-ray pulse from the undulator to the sample have to be taken into account. These are in principle known and should be fixed values for a given synchrotron. In addition, the length of the used cabling has an influence on the phase as well and this changes when the microwave set-up including sample and micro-resonator is physically changed. In a practical experiment the fitted phase $\varphi$ is only a relative number and comprises all the above factors. Therefore, it can only be compared as long as the entire microwave set-up as well as the excitation frequency of the STXM-FMR experiment is not changed. This implies, that it is only possible to compare relative phases within the same sample and not between different samples. In other words, the obtained phase $\varphi=-39^{\circ}\pm$5$^\circ$ is basically meaningless for comparing different sample measured in the STXM-FMR. However,  the relative phase of different measurements using the same parameters and sample upon, e.\,g., the reversal of the helicity of the light can be compared and - according to the XMCD effect - should be 180$^{\circ}$. 

\section{Experimental verification of the contrast behaviour}

Having discussed the small variation of the absorption coefficient during precession with a small opening angle together with the presence of a rather pronounced background signal of the SiN-membrane, it is important to investigate the behavior of the dynamic magnetic contrast upon reversal of the helicity of the light to assure that a true XMCD effect is indeed observed.

\subsection{Contrast reversal with helicity}

\begin{figure}[h!]
\center
\resizebox{1\columnwidth}{!}{\includegraphics{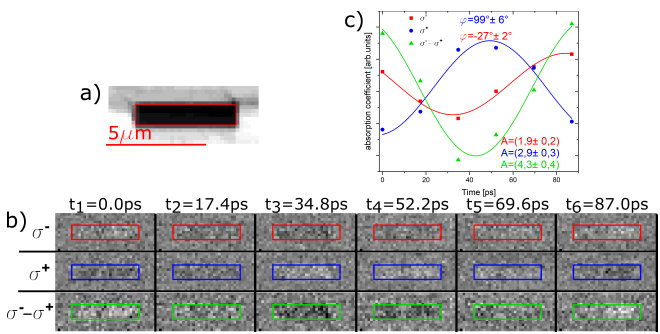}}
\caption{ a) Chemical contrast picture of the Py stripe. In b) two STXM-FMR measurement with different X-ray polarization measured at the Ni-$L_{3}$-edge are shown as well as the difference between the two measurements. c) shows the average transmission intensity of the X-rays through the stripe sample for the three cases shown in b). The averaged data was fitted with a microwave frequency of 9.61\,GHz.}
\label{Fig_VGL_s+_s-}
\end{figure}

\noindent As a first test experiment two STXM-FMR measurements at 9.61\,GHz were done using $\sigma^{+}$ and $\sigma^{-}$ polarized X-rays at the Ni $L_{3}$-edge of the horizontal Py stripe of the T sample shown in Fig.\,\ref{Fig_sa}\,b). In Fig.,\ref{Fig_VGL_s+_s-}\,a) the chemical contrast is shown while the individual 6 phases with $\sigma^{+}$ (blue) and $\sigma^{-}$ (red) light are displayed in Fig.\,\ref{Fig_VGL_s+_s-}\,b), top two rows. All contrast variations in b) are shown on the same scale in order to emphasize the difference in $\Delta \mu \cdot t$ for the different measurements. Fig.\,\ref{Fig_VGL_s+_s-}\,c) collates the averaged dynamic magnetic contrast for all 6 phases derived by averaging over the respective marked areas. The RoIs were identified using the chemical contrast image in Fig.\ref{Fig_VGL_s+_s-}\,a) as indicated by the red box. The same colour scheme was used for the averaged intensities of the two measurements shown in Fig.\ref{Fig_VGL_s+_s-}\,c).\\
As one can clearly see in the individual phase images there is a phase difference of  $-27^{\circ}\pm2^{\circ}$ for $\sigma^-$ and $+99^{\circ}\pm6^{\circ}$ for $\sigma^+$ light. Note, that the value for the phases in Fig.\ \ref{Fig_VGL_ln(ratio)_neweval} and \ref{Fig_VGL_s+_s-} differ because the samples and thus the microwave setup are different. 
The XMCD effect suggests that by reversing the X-ray polarization the relative phase should change by $180^{\circ}$, i.\,e., an ideal reversal of the contrast in all 6 images. However, the relative phase difference between the two measurement is only $127^{\circ}\pm8^{\circ}$, which is significantly smaller. This is most likely due to the experimental constraint that only 6 phases can be resolved because of the X-ray pulse length of 50\,ps, while the time difference between the individual phases is only 17.4\,ps. Therefore, the experimental uncertainty is larger, than the errors from the fitting procedure, especially considering the small overall size of the dynamic contrast change. In turn, one can increase the magnetic contrast by taking the difference between the two experiments with $\sigma^{+}$ and $\sigma^{-}$ light according to usual XMCD experiments. The result is shown in Fig.\ \ref{Fig_VGL_s+_s-}\,b) and c) (green) and it is obvious that the dynamic contrast is enhanced significantly. Nevertheless, as visible in Fig.\,\ref{Fig_VGL_s+_s-}\, c) the amplitude is not increased by a factor of 2 as expected which is due to the non-ideal reversal of the contrast as reflected by the behavior of the relative phases. Nevertheless, this is a first indication that a homogeneous FMR excitation behaves the same way as previously observed spin wave excitations \cite{Bon15}.\\

\subsection{Helicity versus field direction}
\begin{figure}[h!]
\center
\resizebox{1\columnwidth}{!}{\includegraphics{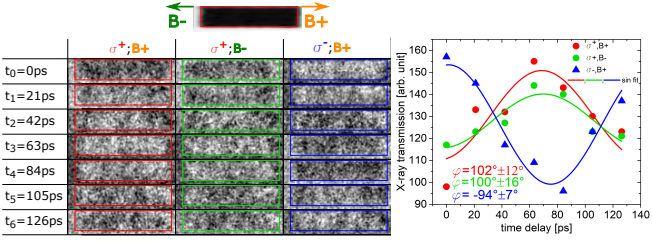}}
\caption{STXM-FMR measurement done at the Maxymus beamline at the Fe-$L_{3}$-edge at B+=60\,mT and B-=-60\,mT and different X-ray polarization $\sigma^+\,,\,\sigma^-$. The applied microwave frequency for all measurements shown in this figure was 6.785\,GHz. The left hand side shows the chemical contrast image together with the different directions for B- and B+. The averaged area is indicated by the red box. Below, the normalized intensity (with respect to the time average state) for the 7 different excitation phases (or delay times) is shown for different field directions and X-ray polarisations. The spatially averaged intensity for each of these boxes can be found on the right hand side with their respective colour coding.}
\label{Fig_Maximus}
\end{figure}
\noindent In Fig.\,\ref{Fig_Maximus} a second control STXM-FMR measurement done at the Maxymus endstation at BESSY II is shown where the STXM-FMR was measured with a slightly increased phase resolution of 7 images for one full precession cycle. In addition to reversing the helicity of the light one helicity was also measured for two different external magnetic field orientations $B_{ext}$, i.\,e., ($\sigma^+,B^+$), ($\sigma^+,B^-$), and ($\sigma^-,B^+$). The relative orientation of $B_{ext}$ is indicated together with the image of the chemical contrast in Fig.\,\ref{Fig_Maximus}\,a), top. The RoI where the contrast is spatially averaged is indicated by the red boxes in all images to ensure that the observed averaged signal only originates from the stripe and does not contain the SiN-membrane background (see above). The time-normalized spatial average intensity for each phase of the 3 different measurements is shown in Fig.\,\ref{Fig_Maximus}\,b)-d). The measurement in a positive magnetic field B+ and circular polarization $\sigma^{+}$ is shown on the left side of Fig.\,\ref{Fig_Maximus}\,a). The resulting averaged normalized X-ray transmission can be found in Fig.\,\ref{Fig_Maximus}\,b), where it was fitted using a sine function with the microwave frequency of 6.785\,GHz. This fit yields a relative phase between the X-ray pulses and the magnetization precession of $102^{\circ}\pm 12^{\circ}$.\\
As mention above the sign of the static XMCD effect reverses when the external magnetic field along an axis of sensitivity is reversed \cite{STO95, DUR09}. At first glance it could be expected that this leads to a contrast reversal for the dynamic magnetic contrast in STXM-FMR as well. However, as can be seen in Fig.\ref{Fig_Maximus}\,a) the contrast does not reverse when the external field is reversed (middle column). This can be explained due to the fact that the STXM-FMR in the present configuration is only sensitive to the transversal dynamic component of the magnetization precession and not the direction of the magnetization itself. Due to the field reversal the magnetization still precesses around the external field with the same phase relation as before. The dynamic magnetic contrast is only dependent on the projection of the dynamic magnetization onto the X-ray k-vector. This projection in turn exhibits a cosine behaviour, thus does not depend on the sense of rotation regarding the X-ray k-vector. This is evidenced by comparing the averaged dynamic magnetic contrast for $\sigma^+,B^+$ in Fig.\,\ref{Fig_Maximus}\,b) and $\sigma^+,B^-$ in c). The resulting relative phase is $102^{\circ}\pm 12^{\circ}$ and $100^{\circ}\pm16^{\circ}$, respectively, i.\,e., identical within error bars. In contrast, comparing $\sigma^{+}$ with $\sigma^{-}$ polarization for $B^+$ in Fig.\,\ref{Fig_Maximus}\,b) and d), respectively, a clear contrast reversal is seen which is reflected by the resulting relative phases of $102^{\circ}\pm 12^{\circ}$ and $-94^{\circ}\pm7^{\circ}$. The resulting phase change is thus $196^{\circ}\pm 19^{\circ}$ which agrees within error bars with the expected value of $180^{\circ}$ for the ideal XMCD effect.

\subsection{Contrast reversal for inhomogeneous excitations}

Until now all measurements were done for a uniform magnetic excitation of the magnetic specimen. There was a drastic difference in conventional XA for areas where only background effects were observed to regions where the dynamic magnetic contrast was measured. Since we have already attributed the oscillating contrast of the background to an interaction of the microwave with the APD, non-linearities of the APD can also have a non-negligible influence on the extracted dynamic magnetic contrast. This is especially relevant for homogeneous excitations of the magnetic specimen. In order to exclude this, an inhomogeneous excitation of the magnetic specimen is the method of choice. 

\begin{figure}[h!]
\center
\resizebox{1\columnwidth}{!}{\includegraphics{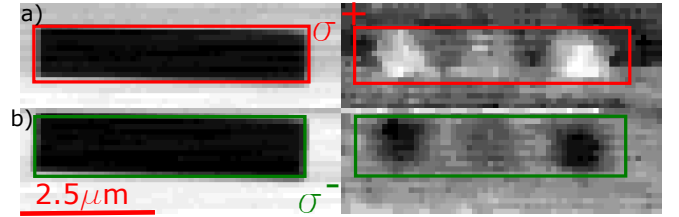}}
\caption{Comparison of the two circular polarizations for an inhomogeneous excitation of a T stripe. The left part shows the chemical contrast pictures for the different polarization measurements respectively. For contrast maximization the opposite phases of the same measurement are subtracted. The red and green boxes indicate the position of the Py stripe (extracted from the chemical contrast) for each of the two measurements to better visualize the excitation.}
\label{Fig_inhomo}
\end{figure}

\noindent In Fig.\,\ref{Fig_inhomo} a STXM-FMR experiment, measured at the SSRL, of an inhomogeneous excitation of the stripe parallel to the external magnetic field of a Py "T"-sample is shown. Details on these types of excitations go beyond the scope of this paper and will be discussed elsewhere; integral FMR measurements together with micromagnetic simulations have already been published \cite{BAN11}. Figure \ref{Fig_inhomo} a) shows the measurement with $\sigma^{+}$ light, whereas in b) the $\sigma^{-}$ case can be seen. On the left-hand side the chemical contrast images are provided while on the right-hand side a single image of the dynamic magnetic contrast is displayed. In order to maximize the contrast, the difference between the same two opposite phases has been taken for both polarizations. An additional smoothing as in Ref.\ \cite{SCH17} has been carried out to better visualize the inhomogeneous excitation. One can clearly see that there are regions with a pronounced magnetic contrast to either end of the stripe while the center shows a much weaker contrast with zero contrast in between. Importantly, the regions of strong magnetic contrast clearly reverse upon reversal of the helicity of the light while in other regions the contrast remains unaffected. Therefore, the contrast reversal is also observable with respect to a non-reversing region where the overall XA does not change, underlining that the contrast mechanism is indeed of magnetic origin.\\

\section{Conclusion}
We have shown a way to correctly separate the quantitative pure dynamic magnetic contrast from the background signal in STXM-FMR. The opening angle of the FMR excitation of Py was determined at the Ni-L$_3$-edge by evaluating the amplitude of the dynamic magnetic contrast yielding an opening angle of 0.15$^{\circ}$ which corresponds well with previously reported values for Co\cite{Bon15}. Furthermore, by switching the polarization of the X-ray photons from $\sigma^{+}$ to $\sigma^{-}$ the dynamic magnetic contrast switches its sign for the STXM-FMR measurement.  However, contrary to static classical XMCD a reversal of the external magnetic field does not change the dynamic magnetic contrast of the STXM-FMR because of the transversal geometry. Enhancement of the signal can be achieved by measuring the STXM-FMR with different polarizations ($\sigma^+$ and $\sigma^-$). Finally, the contrast reversal upon reversal of the helicity was observable for two different STXM-FMR setups, at two different synchrotrons, as well as for an inhomogeneous excitation. This evidences that the contrast in STXM-FMR behaves similar under reversal of the X-ray helicity to the static XMCD effect and one can take advantage from the unique combination of element selectivity and spatio-temporal resolution in future studies of magnetically excited micro- an nano-structures.

\acknowledgments{This research was funded the Austrian Science Foundation (FWF), project No I-3050 as well as the German Research Foundation (DFG), project No OL513/1-1Use of the Stanford Synchrotron Radiation Lightsource, SLAC National Accelerator Laboratory, is supported by the U.S. Department of Energy, Office of Science, Office of Basic Energy Sciences under Contract No. DE-AC02-76SF00515. Part of the measurements were carried out at the MAXYMUS endstation at BESSY II at the Helmholtz-Zentrum Berlin.  We thank HZB for the allocation of synchrotron radiation beamtime.}


\begin{thebibliography}{99}

\bibitem{POO97} Poole, Ch. P.  {\em Electron spin resonance: A Comprehensive Treatise and Experimental Techniques}; Dover Publications Inc. (1997)

\bibitem{NAR05} Narkowicz, R., Suter, D. and Stonies, R. Planar microresonators for EPR experiments. {\em J. Magn. Reson.} {\bf 2005} {\em 175}, 275

\bibitem{BAN11} Banholzer, A., Narkowicz, R., Hassel, C., Meckenstock, R., Stienen, S., Posth, O., Suter, D., Farle, M. and Lindner, J. Visualization of spin dynamics in single nanosized magnetic elements. {\em Nanotechnol.} {\bf 2011}, {\em 22}, 295713 

\bibitem{ROS02} Rosner, B. T. and van der Weide, D. W. High-frequency near-field microscopy. {\em Rev. Sci. Instrum.} {\bf 2002} {\em 73}, 2505 

\bibitem{DEM01} Demokritov, S., Hillebrands, B. and Slavin, A. N. Brillouin light scattering studies of confined spin waves: linear and nonlinear confinement. {\em Phys. Rep.} {\bf 2001} {\em 348}, 441 

\bibitem{VOL04} Volodin, A., Buntinx, D., Brems, S. and Van Haesendonck, C. Piezoresistive detection-based ferromagnetic resonance force microscopy of microfabricated exchange bias systems. {\em Appl. Phys. Lett.} {\bf 2004} {\bf 85}, 5935 

\bibitem{SCH17} Schaffers, T., Meckenstock, R., Spoddig, D., Feggeler, T., Ollefs, K., Sch\"oppner, C., Bonetti, S., Ohldag, H., Farle, M. and Ney, A. The combination of micro-resonators with spatially resolved ferromagnetic resonance. {\em Rev. Sci. Instrum.} {\bf 2017} {\em 88}, 093703

\bibitem{SCH18} Sch\"onke, D., Oelsner, A., Krautscheid, P., Reeve, R. M. and Kl\"aui, M. Development of a scanning electron microscopy with polarization analysis system for magnetic imaging with ns time resolution and phase-sensitive detection. {\em Rev. Sci. Instrum.} {\bf 2018} {\em 89}, 083703

\bibitem{CHE12} Cheng, X. M. and Keavney, D. J. Studies of nanomagnetism using synchrotron-based x-ray photoemission electron microscopy (X-PEEM). {\em Rep. Prog. Phys.} {\bf 2012} {\em 75}, 026501

\bibitem{SCH86} Sch\"utz, G., Wagner, W., Wilhelm, W., Kienle, P., Zeller, R., Frahm, R. and Materlik, G. Absorption of circularly polarized x rays in iron. {\em Phys. Rev. Lett.} {\bf 1986} {\em 58}, 737

\bibitem{STO95} St\"ohr, J. X-ray magnetic circular dichroism spectroscopy of transition metal thin films. {\em J. Electron Spectrosc. Relat. Phenom.} {\bf 1995} {\em 75}, 253

\bibitem{DUR09} D\"urr, H. A., Eim\"uller, T. Elmers, H.-J., Eisebitt, S., Farle, M., Kuch, W., Matthes, F., Martins, M., Mertins, H. C., Oppeneer, P. M., Plucinski, L., Schneider, C. M., Wende, H., Wurth W. and Zabel, H. A Closer Look Into Magnetism: Opportunities With Synchrotron Radiation. {\em IEEE Trans. Magnet.} {\bf 2009} {\em 45}, 15

\bibitem{OMS15} Ollefs, K., Meckenstock, R., Spoddig, D., R\"omer, F. M., Hassel, C., Sch\"oppner, C., Ney, V., Farle, M. and Ney, A. Toward broad-band x-ray detected ferromagnetic resonance in longitudinal geometry. {\em J. Appl. Phys.} {\bf 2015} {\em 117}, 223906

\bibitem{PUZ05} Puzic, A., Van Waeyenberge, V., Chou, K. W., Fischer, P., Stoll, H., Sch\"utz, G., Tyliszczak, T., Rott, K., Br\"uckl, H., Reiss, G., Neudecker, I., Haug, T., Buess M. and Back, C. H. Spatially resolved ferromagnetic resonance: Imaging of ferromagnetic eigenmodes. {\em J. Appl. Phys.} {\bf 2005} {\em 97}, 10E704

\bibitem{Bon15} Bonetti, S., Kukreja, R., Chen, Z., Spoddig, D., Ollefs, K., Sch\"oppner, C., Meckenstock, R., Ney, A., Pinto, J., Houanche, R., Frisch, J., St\"ohr, J., D\"urr, H. and Ohldag, H. Microwave soft x-ray microscopy for nanoscale magnetization dynamics in the 5-10 GHz frequency range. {\em Rev. Sci. Instrum.} {\bf 2015} {\em 86}, 093703

\bibitem{WEI14} Weigand, M., Realization of a new Magnetic Scanning X-ray Microscope and Investigation of Landau Structures under Pulsed Field Excitation. (Dissertation), Cuvillier Verlag, G\"ottingen (2014)

\bibitem{STE13} Stein, F. U., Bocklage, L., Weigand, M. and Meier, G. Time-resolved imaging of nonlinear magnetic domain-wall dynamics in ferromagnetic nanowires. {\em Sci. Rep.} {\bf 2013} {\bf  3}, 1737

\bibitem{WIN16} Wintz, S., Tiberkevich, V., Weigand, M., Raabe, J., Linder, J., Erbe, A., Slavin A. and Fassbender, Magnetic vortex cores as tunable spin-wave emitters. {\em Nat. Nanotech.} {\bf 2016} {\em 11} 948

\bibitem{FEG19} Feggeler, T. et al. Direct visualization of dynamic magnetic coupling in a Co/Py double layer with ps and nm resolution. arXiv:1905.06772 (2019)

\bibitem{KUK15} Kukreja, R., Bonetti, S., Chen, Z., Backes, D., Acremann, Y., Katine, J. A., Kent, A. D., D\"urr, H.,A., Ohldag, H. and St\"ohr, J. X-ray Detection of Transient Magnetic Moments Induced by a Spin Current in Cu. {\em Phys. Rev. Lett.} {\bf 2015} {\em 115} 096601

\bibitem{SWI62} Swinehart, D. F. The Beer-Lambert Law. {\em J. Chem. Educ.} {\bf 1962} {\em 39}, 333

\end{thebibliography}
\end{document}